\documentclass[conference]{IEEEtran}
\ifCLASSINFOpdf
  \usepackage[pdftex]{graphicx}
  \graphicspath{{images/}}
  \DeclareGraphicsExtensions{.pdf}
\else
\fi
\usepackage[caption=false,font=normalsize,labelfont=sf,textfont=sf]{subfig}

\usepackage{flushend} 
\usepackage{hyperref} 
\usepackage{amsthm} 
\usepackage{amsmath} 
\usepackage{mathtools} 
\usepackage{amssymb} 
\newtheorem{definition}{Definition}

\usepackage{glossaries}

\newacronym{mde}{MDE}{Model-driven engineering}
\newacronym{bpmn}{BPMN}{Business Process Modeling Notation}
\newacronym{ct}{CT}{category theory}
\newacronym{uml}{UML}{Unified Modeling Language}
\newacronym{dsl}{DSL}{Domain Specific Language}
\newacronym{csp}{CSP}{Communication Sequential Processes}

\begin{document}
\title{Towards behavioral consistency \\ in heterogeneous modeling scenarios}

\author{\IEEEauthorblockN{Tim Kräuter}
\IEEEauthorblockA{Høgskulen på Vestlandet\\
Bergen, Norway\\
Email: tkra@hvl.no}}

\maketitle

\begin{abstract}
Behavioral models play an essential role in \gls{mde}.
Keeping inter-related behavioral models consistent is critical to use them successfully in \gls{mde}. 
However, consistency checking for behavioral models, especially in a heterogeneous scenario, is limited.

We propose a methodology to integrate heterogeneous behavioral models to achieve consistency checking in broader scenarios.
It is based on aligning the respective behavioral metamodels by defining possible inter-model relations which carry behavioral meaning.
Converting the models and their relations to a behavioral formalism enables analysis of global behavioral consistency using model-checking. 
\end{abstract}


\IEEEpeerreviewmaketitle

\section{Problem}
A significant motivation for \glsfirst{mde} is to handle the increasing complexity of software systems by a clear separation of concerns \cite{franceModeldrivenDevelopmentComplex2007}.
In the multi-view modeling approach, a set of models is developed for the different aspects of a system.
Consequently, it is likely that separate groups of people work independently on different parts of the system.
However, this separation of concerns causes problems because models must be kept consistent, i.e., they should not contain contradicting information \cite{cicchettiMultiviewApproachesSoftware2019}.
Without this inter-model consistency, \gls{mde} cannot deliver on the promised productivity increase and error reduction \cite{brambillaModeldrivenSoftwareEngineering2017}.

Inter-model consistency is especially problematic when the used models are heterogeneous:
Firstly, structural models such as class diagrams and entity-relationship diagrams exist.
Secondly, behavioral diagrams are used to describe the dynamics of a system.
Widely used diagrams or formalisms for specifying behavior are state machines, activity diagrams, Petri nets, and process algebras.

Inter-model consistency for structural models has been researched extensively.
One promising approach is model weaving \cite{bezivinCanonicalSchemeModel2006}, which establishes links between models with the goal of automatic consistency checking.
Typically, these inter-relations focus on structural aspects, such as identity, usage, dependency, and refinement \cite{feldmannManagingIntermodelInconsistencies2019, torresSystematicLiteratureReview2020}.
They are called \textit{correspondences} on the metamodel level and \textit{commonalities} on the model level \cite{stunkelMultipleModelSynchronization2020, klareCommonalitiesPreservingConsistency2019}.
Inter-model constraints can then be defined and automatically be checked.
This works by either merging individual models to a global model using commonalities \cite{stunkelMultimodelCorrespondenceIntermodel2018} or establishing a comprehensive view of the global model \cite{stunkelMultipleModelSynchronization2020}.

However, there is no general approach to check global behavioral consistency on a collection of models, sometimes called semantic consistency.
Behavioral models describing interacting parts of a system should be checked for deadlocks, live-locks, and other system-specific requirements.
This is generally not problematic if the models conform to the same formalism or modeling language.
Nevertheless, in a heterogeneous case, one still wants to define interactions between the models and check behavioral consistency.

Using only one formalism or modeling language for behavioral models is not feasible since one wants to use the most suitable formalism or modeling language in each situation.
The modeling formalisms used in each case depend on the system requirements, existing software landscape, and the knowledge and preferences of the responsible developers. 

There can also be consistency rules spanning structural and behavioral models.
For example, messages to objects in a \gls{uml} sequence diagram must match the corresponding classes methods in a class diagram \cite{egyedFixingInconsistenciesUML2007}.
We will focus on behavioral consistency before addressing consistency between structural and behavioral models in future work. 

\section{Related work} \label{sec:related_work}
\cite{engelsMethodologySpecifyingAnalyzing2001} investigates the consistency of object-oriented behavioral models formulated as capsule statecharts in \gls{uml}-RT.
Consequently, they deal with a homogeneous modeling environment where all behavioral models are developed in the same modeling language.
The authors use the term semantic consistency instead of behavioral consistency.
To check behavioral consistency, they map statechart models to \gls{csp}.
Using the well-defined semantics of \gls{csp}, they validate deadlock freeness and the processes compliance to a previously defined communication protocol.
Their proposed general methodology to analyze consistency is to find a suitable semantic domain and map the models into it.
The semantic domain can then be exploited to check the consistency.

Based on this methodology, consistency checking for sequence diagrams and statecharts was developed in \cite{kusterExplicitBehavioralConsistency2003}.
The consistency requirement claims that all possible interactions specified within sequence diagrams for each class should be possible regarding the behavior of that class defined in a statechart.
As a semantic domain, the authors have successfully used \gls{csp} again.

Besides process algebras such as \gls{csp}, different types of Petri nets are used for behavioral consistency checking.
In \cite{yaoConsistencyCheckingUML2006}, Petri nets were used to check consistency for sequence diagrams and statecharts.
\cite{cunhaFormalVerificationUML2011} analyses sequence diagrams in the context of embedded systems using a transformation to Petri nets.
Consistency for \gls{uml} activity diagrams was checked using Petri nets in \cite{thierry-miegUMLBehavioralConsistency2008}.

In practice, one could encounter a situation where the combination of state machines, process algebras, Petri nets, and activity diagrams is desired\footnote{The merger of two companies and their IT-infrastructures could for example cause heterogeneity in the used modeling languages.}.
It is possible to check behavioral consistency using our proposed approach even in heterogeneous modeling scenarios.
This leads to more freedom of choice regarding behavioral modeling and model combinations.
In addition, behavioral models of higher quality can be achieved, which lay the foundation for successful and effective \gls{mde}.

Approaches that cover heterogeneous scenarios and allow for model simulation, which can be used for behavioral consistency verification, are Ptolemy \cite{ekerTamingHeterogeneityPtolemy2003} and GEMOC studio \cite{deantoniModelingBehavioralSemantics2016}.
However, they lack formal semantics, or in the case of Ptolemy, formal semantics have been developed afterwards \cite{tripakisModularFormalSemantics2013}.

\section{Proposed solution}
\begin{figure}[h]
    \centering
    \includegraphics[width=3.4in]{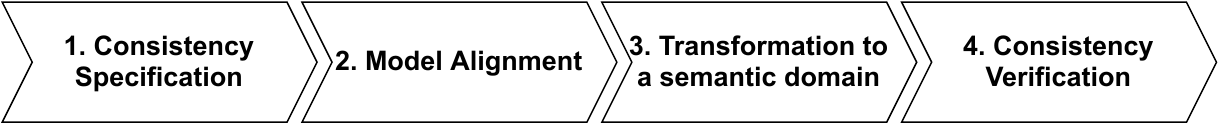}
    \caption{Behavioral Consistency Management Process}
    \label{fig:consistency_process}
\end{figure}

The proposed solution builds on the methodology presented in \cite{engelsMethodologySpecifyingAnalyzing2001} but introduces a new step inspired by the multi-model consistency management process for structural models defined in \cite{stunkelMultipleModelSynchronization2020}.
The proposed solution can be summarized by the four-step process in \autoref{fig:consistency_process}.
We will explain it in detail now.

A potential consistency problem for a set of behavioral models triggers the consistency management process.
The first step of the process begins with a problem discussion and an informal problem statement.
The step concludes by formalizing the consistency problem.
To formalize the problem, one should add atomic propositions to the states of the individual models with the problem statement in mind.
Then, we can calculate the disjoint union of all atomic propositions and use this set of propositions to formalize our constraint in a temporal logic as if we had one model describing the global system.

To check inter-model consistency for structural models, one defines inter-relations between models on the metamodel level and the model level to describe overlaps in information \cite{stunkelMultipleModelSynchronization2020, elhamlaouiAlignmentViewpointHeterogeneous2016}.
We propose a similar approach for behavioral models based on inter-model relations.
The inter-model relations on the metamodel level define how behavioral models can coordinate.
Thus, we call them \textit{coordinations}.
The inter-model relations on the model level should state how models interact, which leads to the composite system's behavior.
Consequently, we call them \textit{interactions}.

The second step is to define inter-model relations, i.e., coordinations and interactions.
Firstly, one has to define coordinations that align the metamodels.
For this approach to work, there must be a metamodel for each used model.
However, developing a metamodel if it is not already present is a one-time expense.
Similarly, metamodel alignment must be only done once.
In \autoref{sec:currentStatus}, we will align the metamodels for finite state machines and Petri nets.
In the future, other formalisms can be integrated similarly.
Secondly, one must align the models by defining interactions between them.
For example, the synchronization of a transition in a state machine with a transition in a Petri net could be specified.
\textit{Model alignment} is also an essential activity in the multi-model consistency management process for structural models \cite{stunkelMultipleModelSynchronization2020}.

The third step in the process is to transform the models into a suitable semantic domain respecting their interactions.
By including the inter-model relations in the transformation, we can deal with a heterogeneous modeling scenario compared to the original methodology in \cite{engelsMethodologySpecifyingAnalyzing2001}.
A feasible approach is to translate each model individually and then combine them according to the defined interactions.

We will use graph grammars as a behavioral formalism for the model execution (see \autoref{sec:currentStatus}) for the following reasons.
Many behavioral formalisms have been described by graph grammars, for example, state machines \cite{kuskeFormalSemanticsUML2001, varroFormalSemanticsUML2002}, Petri nets \cite{ehrigGraphGrammarsPetri2004}, the $\pi$-calculus \cite{gadducciGraphRewritingPcalculus2007}, and workflow models \cite{rutleMetamodellingApproachBehavioural2012}.
Using graph grammars, one stays on a higher level of abstraction than, for example, transition systems as a semantic domain.
This has the advantage that the state space generated by graph grammars is better understandable.
Counterexamples for consistency constraints should be easier to understand because we stay closer to the original modeling formalisms.
Being on a higher level makes it easier to implement different communication variants such as synchronous and asynchronous message passing.
In addition, structural models are formalized as graphs or graph-like structures \cite{stunkelMultipleModelSynchronization2020}, which can lead to an integration of behavioral and structural models using graph transformations in the future. 
However, the semantic domain is interchangeable if there is an interpretation for the given models and their interactions.

The fourth step in the process is concerned with verifying the specified behavioral consistency.
The semantic domain should generate a state space for the overall system that is related to the original models.
This should allow us to attach the atomic propositions from the individual models to the generated state space.
Thus, one gets a Kripke structure (state space and atomic propositions \cite{clarkeHandbookModelChecking2018}) to check behavioral constraints.
Besides checking general properties such as deadlocks, the model-specific constraints defined in step one should be checked.
We aim to reuse existing model checkers to evaluate the defined constraints on the resulting Kripke structure.

In the best case, the specification of constraints is independent of the choice of the semantic domain, such that the behavioral constraints can be checked automatically.
Generally, user input is also possible in this step if the chosen semantic domain does not precisely match the previously defined constraints.
However, this is not desirable since the semantic domain should be mostly hidden from the user.

\section{Plan for evaluation and validation}
To evaluate our methodology, we plan to implement a tool for behavioral consistency checking.
It should come with a textual \gls{dsl} for metamodel and model alignment, automatic transformations to a semantic domain, and the semantic domain's implementation, including consistency verification.
For graph grammars, one can use tools like GROOVE \cite{ghamarianModellingAnalysisUsing2012, rensinkGROOVESimulatorTool2004}, or Verigraph \cite{costaVerigraphSystemSpecification2016} to generate a state space from a graph grammar.

The developed tool serves as a basis for evaluation and validation.
At first, we plan to validate if the tool sufficiently covers frequent combinations of behavioral models found in the literature and industry.
Afterward, we will evaluate the tool based on its performance and scalability by either increasing the number of behavioral models and their size or the computing resources available.
Finally, we will compare our approach to the approaches taken by GEMOC studio \cite{deantoniModelingBehavioralSemantics2016} and Ptolemy \cite{ekerTamingHeterogeneityPtolemy2003}.
The comparison will be based on the evaluation criteria mentioned before and the supported behavioral model heterogeneity.

\section{Expected contributions}
We plan to make the following contributions:
\begin{itemize}
    \item A new methodology for behavioral consistency checking in heterogeneous cases.
    \item Metamodels and metamodel alignments for common behavioral modeling formalisms.
    \item A semantic domain for advanced behavioral consistency checking.
\end{itemize}
Besides these theoretical contributions, we plan to develop a modeling tool for metamodel and model alignment and behavioral consistency checking following the proposed methodology.

\section{Current status} \label{sec:currentStatus}
Our work on behavioral consistency has just started, such that we do not have a finished implementation of the following theoretical aspects yet.
So far, finite state machines and Petri nets have been considered as behavioral models/formalisms.

\begin{figure*}[h]
    \centering
    \includegraphics[width=5in]{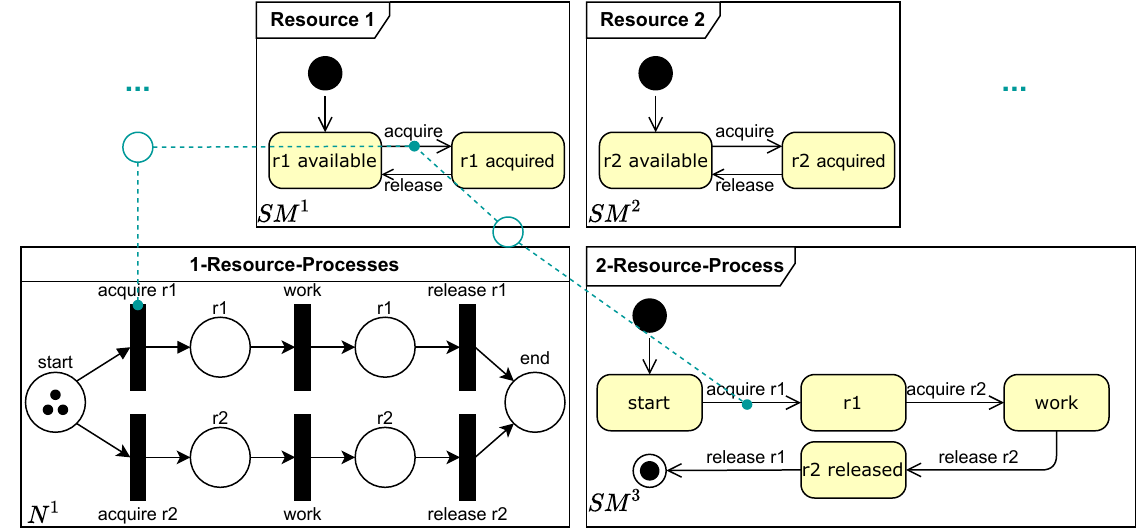}
    \caption{Example models $SM^1$, $SM^2$, $SM^3$, $N^1$, and their interactions}
    \label{fig:running_example}
\end{figure*}

We will use the example models depicted in \autoref{fig:running_example} to explain the current status of our work by following the proposed methodology.
\autoref{fig:running_example} depicts three state machines, $SM^1$-$SM^3$, and a Petri net $N^1$ (ignore the cyan connections for now).
As a convention, Petri net edges have the weight one if not explicitly stated otherwise.
The state machines $SM^1$ and $SM^2$ represent two resources that can be acquired and released by the processes defined by $SM^3$ and $N^1$.
$SM^3$ acquires both resources, does some work, and then releases them, while the three processes modeled in the Petri net $N^1$ only need one of the two resources to do their work.
What each process is doing with the resources is kept abstract.
However, in practice, we often face situations in which multiple processes are competing for the same resources.

One could have used Petri nets to model the state machines in \autoref{fig:running_example}.
Nevertheless, we have used state machines to model most processes of the running example since the extra features of Petri nets were not needed.
In this case, the difference between a Petri net model and a state machine model for a resource is tiny, but in general, one formalism can be better suited for a given aspect of a system.

A reasonable consistency requirement for the overall system resulting from executing the models in parallel is that all processes always finish their execution.
In addition, each resource is accessed by at most one process at a time.
Thus, we formulate the following constraints:
\begin{enumerate}
    \item \textbf{Successful termination:} The state where both resources are available while $SM^3$ is in its final state and $N^1$ has three tokens in the place \textit{end} should eventually be reached.
    \item \textbf{Proper resource one access:} There can be at most one token in the places \textit{r1} while $SM^3$ is in the state \textit{start} or \textit{end}.
    If $SM^3$ is not in the state \textit{start} or \textit{end}, there must be no token in the places \textit{r1}.
    \item \textbf{Proper resource two access:} There can be at most one token in the places \textit{r2} while $SM^3$ is not in the state \textit{work}.
    If $SM^3$ is in the state \textit{work}, there must be no token in the places \textit{r2}.
\end{enumerate}

The constraints can be formalized using temporal logic by assigning appropriate atomic propositions to the mentioned states.
The first constraint can be seen as a liveness property, while the last two constraints describe safety properties.

To check our constraints, we need to define finite state machines and Petri nets, state their respective metamodels, and align them.
After that, we can align the models of our running example depicted in \autoref{fig:running_example}.

\begin{definition}[Finite state machine] \label{def:fsm}
    A finite state machine $M=(S, \Sigma, \delta, s_0, F)$ consists of a set of states $S$, a finite alphabet $\Sigma$, a state transition relation $\delta \subseteq S \times \Sigma \times S$, an initial state $s_0 \in S$ and a set of final states $F \subseteq S$ \cite{kunzeBehaviouralModelsModelling2016}. 
\end{definition}

\begin{figure}[h]
    \centering
    \includegraphics[width=2.3in]{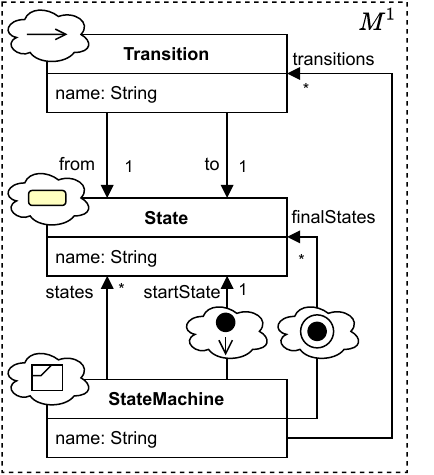}
    \caption{Finite state machine metamodel $M^1$}
    \label{fig:fsm_metamodel}
\end{figure}

\autoref{fig:fsm_metamodel} depicts the metamodel $M^1$ for finite state machines.
In the remainder of this paper, we will use the term state machines to refer to finite state machines.
In addition to \autoref{def:fsm}, we added names to states and state machines.
The finite alphabet $\Sigma$ is not made explicit but can be derived by the transition names of a given state machine.
The clouds in \autoref{fig:fsm_metamodel} depict concrete syntax elements inspired by \gls{uml} statecharts.
The concrete syntax elements for state machines and the one following for Petri nets are already used in \autoref{fig:running_example}.

\begin{definition}[Petri net] \label{def:pn}
    A Petri net $N=(P,T,F, \omega)$ consists of a finite set of places $P$, a non-empty finite set of transitions $T$ ($T \cap P = \emptyset $), a flow relation $F \subseteq (P \times T) \cup (T \times P)$, and a weighting function $\omega: F \to \mathbb{N}$ \cite{kunzeBehaviouralModelsModelling2016}. 
\end{definition}

\begin{figure}[h]
    \centering
    \includegraphics[width=3.4in]{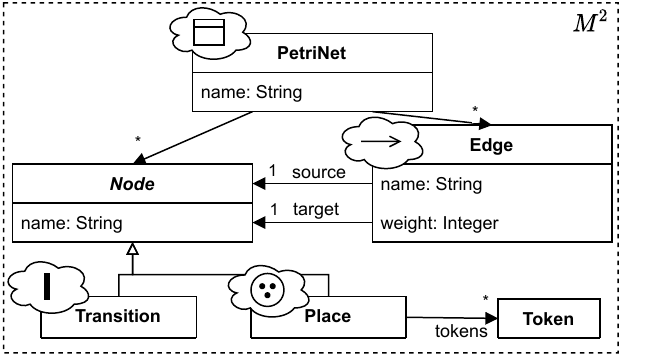}
    \caption{Petri net metamodel $M^2$}
    \label{fig:petri_net_metamodel}
\end{figure}

\autoref{fig:petri_net_metamodel} illustrates the metamodel $M^2$ for Petri nets\footnote{Formally edges representing elements of the flow relation can have transitions or places as source and target which does not conform to \autoref{def:pn}.
We will ignore this issue for now, since we do not want to introduce constraints to the metamodel.}.
In addition to \autoref{def:pn}, places have a set of tokens since we want to model states of Petri nets, which are token distributions.
We again added names to transitions, places, edges, and the Petri net itself.
The concrete syntax for the metamodel is also given, where edge weights should be written above the arrow depicting the edge.
The default weight is one if no weight is specifically stated.

\begin{figure}[h]
    \centering
    \includegraphics[width=3.48in]{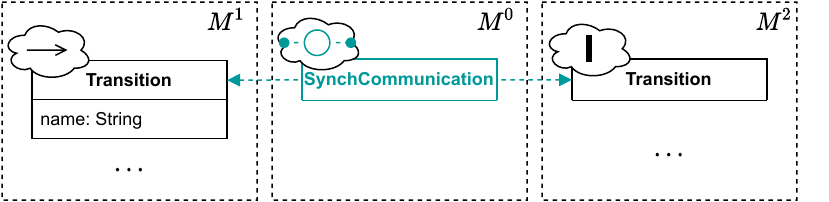}
    \caption{Alignment of the metamodels $M^1$ and $M^2$}
    \label{fig:fsm_pn_alignment}
\end{figure}

Aligning the metamodels for finite state machines and Petri nets is straightforward since both formalisms rely on transitions to change the application state.
Consequently, transitions are aligned, as shown in \autoref{fig:fsm_pn_alignment}.
The inter-model relation relating the two is called synchronous communication and should be interpreted as two transitions firing simultaneously, i.e., handshaking in a semantic domain.
One could also add an inter-model relation for asynchronous communication between transitions.

It is sufficient to interpret \autoref{fig:fsm_pn_alignment} as defining the existence of a set of pairs of transitions $SynchCom \subseteq \delta \times T$.
A different formalization suitable for more complex model alignment is the comprehensive system construction \cite{stunkelMultipleModelSynchronization2020}.

The cyan connections in \autoref{fig:running_example} suggest the model alignment for our running example.
All \textit{acquire r1}/\textit{acquire r2} transitions are synchronized with the \textit{acquire} transition for the resource 1/2.
Formally we obtain four sets of synchronizations since we have two models synchronizing with two resources.
Aligning the state machine metamodel with itself (formally needed for the alignment of $SM^2$ and $SM^3$) is not explicitly shown but is analogous to the alignment shown in \autoref{fig:fsm_pn_alignment} for Petri nets.
The model alignment states how the models interact with each other, which is crucial for the state space generation in the chosen semantic domain.

We will choose graph grammars as a semantic domain for the reasons stated earlier, but one could also choose a different domain or switch in the future.
Our approach is to transform each model into a graph grammar and combine the resulting grammars to realize the defined coordination between the models.
The semantic domain of graph grammars is defined as follows.

\begin{definition}[Graph grammar] \label{def:graphGrammar}
A graph grammar $GG=(S, P)$ consists of a start graph $S$ and a set of production rules $P$ \cite{ehrigGraphGrammarsPetri2004}. 
\end{definition}
\begin{definition}[Production rule] \label{def:productionRule}
A production rule $P= L \overset{l}{\leftarrow} K \overset{r}{\to} R$ consists of graphs L, K, R, and injective graph morphisms $l: K \to L$ and $r: K \to R$ \cite{ehrigGraphGrammarsPetri2004}. 
\end{definition}

Informally speaking, elements in $R$ but not in $L$ are added by a rule, while elements in $L$ and $R$ are preserved.
A rule deletes elements that are in $L$ but not in $R$.

The transformation from Petri nets to graph grammars is based on the semantics defined in \cite{ehrigGraphGrammarsPetri2004}.
We assume that the token distribution in a Petri-net model $pn$ corresponds to its initial state.
Thus, the start graph is given by the graph, which only contains the places and tokens from $pn$.
Each transition is transformed to a production rule which removes tokens from the incoming places of the transition and adds tokens to the outgoing places of the transition according to the defined weights.
For the Petri net $N^1$, the resulting graph grammar will have six rules, where each rule removes one token from the previous place and adds one token to the following place.

The initial state of a graph grammar representing a state machine is given by the graph only containing its start state.
Again, each transition is transformed to a production rule that removes the transition's source state and adds the transition's target state.
Consequently, during the execution of a state machine, the corresponding graph will always only contain the current state.

Combining a set of graph grammars for state machines and Petri nets is given by the union of start states and the merge of production rules.
All production rules without synchronization carry over unchanged.
Every pair in a \textit{SynchCom} set leads to one rule where the three graphs of the rule are the component-wise union of the two underlying rules.

\begin{figure}[h]
    \centering
    \includegraphics[width=2.6in]{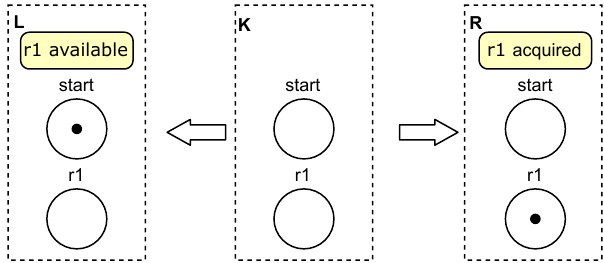}
    \caption{Resulting rule (acquire, acquire r1) for \autoref{fig:running_example}}
    \label{fig:combined_rule}
\end{figure}

In our running example, the synchronization (cyan connection) between $SM^1$ and $N^1$ leads to the rule shown in \autoref{fig:combined_rule}.
However, since we have four synchronization sets, we also get a joint rule where \textit{acquire} of $SM^1$ synchronizes with \textit{acquire r1} in $SM^3$.
This means resource one either synchronizes with $N^1$ or $SM^3$, but never both.

The state space generated by the final graph grammar is essentially a combination of states of the individual state spaces.
Consequently, we can carry over the atomic propositions from the individual models and even check local constraints for individual models in the global state space.
More interestingly, one can check the inter-model behavioral constraints defined earlier using the union of atomic propositions.

For our running example, we would generate the state space from the previously obtained graph grammar and attach the set of defined atomic propositions.
Afterwards, we can check the consistency constraints formulated in the first step of the consistency management process.
Eventual counterexamples to our constraints should be understandable since the graph grammar directly operates on instances of the defined metamodels.
We checked the behavioral consistency for our running example using Groove\footnote{\url{https://github.com/timKraeuter/MODELS-2021-Doctoral-Symposium/tree/main/example_implementation_groove}.}.

We plan to finish the theoretical foundations in the next few months by defining executable algorithms for the transformation of state machines and Petri nets to the semantic domain.
Afterwards, during the last month of this year, we plan to implement a prototype to automate the transformation to the semantic domain, including consistency verification.
In parallel, we integrate the $\pi$-calculus in the semantics domain of graph grammars inspired by \cite{gadducciGraphRewritingPcalculus2007}.
Other popular behavioral models such as \gls{uml} activity diagrams and \gls{bpmn} models should be investigated in the near future.

\section{Conclusion}
We proposed a methodology for handling heterogeneous behavioral model consistency.
To the best of our knowledge, current approaches only offer limited consistency checking in a heterogeneous scenario.
To cope with heterogeneity, metamodel and model alignment are proposed inspired by the model alignment of structural models.

Furthermore, we have proposed graph grammars as a suitable semantic domain for behavioral consistency checking.
Finite state machines and Petri nets, as two fundamental behavioral formalisms, have been encoded in the graph grammar domain.
Their metamodels have been aligned, and the intended synchronization semantics were highlighted using a running example.

Possible future work is adding asynchronous communication, synchronizing multiple Petri net transitions with one state machine transition, or synchronizing more than two transitions from different models with each other.
In addition, more behavioral modeling formalisms must be included to allow for more heterogeneity.
One could also aim to include continuous-time models for more diverse modeling scenarios.

\section*{Acknowledgment}
The author would like to thank Harald König, Adrian Rutle, and Yngve Lamo for fruitful discussions about the topic and the anonymous reviewers for their valuable comments and helpful suggestions.

\bibliographystyle{IEEEtran}
\bibliography{bib}

\begin{thebibliography}{10}
\providecommand{\url}[1]{#1}
\csname url@samestyle\endcsname
\providecommand{\newblock}{\relax}
\providecommand{\bibinfo}[2]{#2}
\providecommand{\BIBentrySTDinterwordspacing}{\spaceskip=0pt\relax}
\providecommand{\BIBentryALTinterwordstretchfactor}{4}
\providecommand{\BIBentryALTinterwordspacing}{\spaceskip=\fontdimen2\font plus
\BIBentryALTinterwordstretchfactor\fontdimen3\font minus
  \fontdimen4\font\relax}
\providecommand{\BIBforeignlanguage}[2]{{%
\expandafter\ifx\csname l@#1\endcsname\relax
\typeout{** WARNING: IEEEtran.bst: No hyphenation pattern has been}%
\typeout{** loaded for the language `#1'. Using the pattern for}%
\typeout{** the default language instead.}%
\else
\language=\csname l@#1\endcsname
\fi
#2}}
\providecommand{\BIBdecl}{\relax}
\BIBdecl

\bibitem{franceModeldrivenDevelopmentComplex2007}
R.~France and B.~Rumpe, ``\BIBforeignlanguage{en}{Model-driven {{Development}}
  of {{Complex Software}}: {{A Research Roadmap}}},'' in
  \emph{\BIBforeignlanguage{en}{Future of {{Software Engineering}} ({{FOSE}}
  '07)}}.\hskip 1em plus 0.5em minus 0.4em\relax {Minneapolis, MN, USA}:
  {IEEE}, May 2007, pp. 37--54.

\bibitem{cicchettiMultiviewApproachesSoftware2019}
A.~Cicchetti, F.~Ciccozzi, and A.~Pierantonio,
  ``\BIBforeignlanguage{en}{Multi-view approaches for software and system
  modelling: A systematic literature review},''
  \emph{\BIBforeignlanguage{en}{Software and Systems Modeling}}, vol.~18,
  no.~6, pp. 3207--3233, Dec. 2019.

\bibitem{brambillaModeldrivenSoftwareEngineering2017}
M.~Brambilla, J.~Cabot, and M.~Wimmer,
  \emph{\BIBforeignlanguage{en}{Model-Driven Software Engineering in
  Practice}}, 2nd~ed., ser. Synthesis Lectures on Software Engineering.\hskip
  1em plus 0.5em minus 0.4em\relax {San Rafael, Calif.}: {Morgan \& Claypool
  Publishers}, 2017, no.~4.

\bibitem{bezivinCanonicalSchemeModel2006}
J.~B{\'e}zivin, S.~Bouzitouna, M.~D. Del~Fabro, M.-P. Gervais, F.~Jouault,
  D.~Kolovos, I.~Kurtev, and R.~F. Paige, ``\BIBforeignlanguage{en}{A
  {{Canonical Scheme}} for {{Model Composition}}},'' in
  \emph{\BIBforeignlanguage{en}{Model {{Driven Architecture}} \textendash{}
  {{Foundations}} and {{Applications}}}}, D.~Hutchison, T.~Kanade, J.~Kittler,
  J.~M. Kleinberg, F.~Mattern, J.~C. Mitchell, M.~Naor, O.~Nierstrasz,
  C.~Pandu~Rangan, B.~Steffen, M.~Sudan, D.~Terzopoulos, D.~Tygar, M.~Y. Vardi,
  G.~Weikum, A.~Rensink, and J.~Warmer, Eds.\hskip 1em plus 0.5em minus
  0.4em\relax {Berlin, Heidelberg}: {Springer Berlin Heidelberg}, 2006, vol.
  4066, pp. 346--360.

\bibitem{feldmannManagingIntermodelInconsistencies2019}
S.~Feldmann, K.~Kernschmidt, M.~Wimmer, and B.~{Vogel-Heuser},
  ``\BIBforeignlanguage{en}{Managing inter-model inconsistencies in model-based
  systems engineering: {{Application}} in automated production systems
  engineering},'' \emph{\BIBforeignlanguage{en}{Journal of Systems and
  Software}}, vol. 153, pp. 105--134, Jul. 2019.

\bibitem{torresSystematicLiteratureReview2020}
W.~Torres, M.~G.~J. {van den Brand}, and A.~Serebrenik,
  ``\BIBforeignlanguage{en}{A systematic literature review of cross-domain
  model consistency checking by model management tools},''
  \emph{\BIBforeignlanguage{en}{Software and Systems Modeling}}, Oct. 2020.

\bibitem{stunkelMultipleModelSynchronization2020}
P.~St{\"u}nkel, H.~K{\"o}nig, Y.~Lamo, and A.~Rutle,
  ``\BIBforeignlanguage{en}{Towards {{Multiple Model Synchronization}} with
  {{Comprehensive Systems}}},'' in \emph{\BIBforeignlanguage{en}{Fundamental
  {{Approaches}} to {{Software Engineering}}}}, H.~Wehrheim and J.~Cabot,
  Eds.\hskip 1em plus 0.5em minus 0.4em\relax {Cham}: {Springer International
  Publishing}, 2020, vol. 12076, pp. 335--356.

\bibitem{klareCommonalitiesPreservingConsistency2019}
H.~Klare and J.~Gleitze, ``\BIBforeignlanguage{en}{Commonalities for
  {{Preserving Consistency}} of {{Multiple Models}}},'' in
  \emph{\BIBforeignlanguage{en}{2019 {{ACM}}/{{IEEE}} 22nd {{International
  Conference}} on {{Model Driven Engineering Languages}} and {{Systems
  Companion}} ({{MODELS}}-{{C}})}}.\hskip 1em plus 0.5em minus 0.4em\relax
  {Munich, Germany}: {IEEE}, Sep. 2019, pp. 371--378.

\bibitem{stunkelMultimodelCorrespondenceIntermodel2018}
P.~St{\"u}nkel, H.~K{\"o}nig, Y.~Lamo, and A.~Rutle,
  ``\BIBforeignlanguage{en}{Multimodel correspondence through inter-model
  constraints},'' in \emph{\BIBforeignlanguage{en}{Conference {{Companion}} of
  the 2nd {{International Conference}} on {{Art}}, {{Science}}, and
  {{Engineering}} of {{Programming}} - {{Programming}}'18
  {{Companion}}}}.\hskip 1em plus 0.5em minus 0.4em\relax {Nice, France}: {ACM
  Press}, 2018, pp. 9--17.

\bibitem{egyedFixingInconsistenciesUML2007}
A.~Egyed, ``\BIBforeignlanguage{en}{Fixing {{Inconsistencies}} in {{UML Design
  Models}}},'' in \emph{\BIBforeignlanguage{en}{29th {{International
  Conference}} on {{Software Engineering}} ({{ICSE}}'07)}}.\hskip 1em plus
  0.5em minus 0.4em\relax {Minneapolis, MN, USA}: {IEEE}, May 2007, pp.
  292--301.

\bibitem{engelsMethodologySpecifyingAnalyzing2001}
G.~Engels, J.~M. K{\"u}ster, R.~Heckel, and L.~Groenewegen,
  ``\BIBforeignlanguage{en}{A methodology for specifying and analyzing
  consistency of object-oriented behavioral models},''
  \emph{\BIBforeignlanguage{en}{ACM SIGSOFT Software Engineering Notes}},
  vol.~26, no.~5, pp. 186--195, Sep. 2001.

\bibitem{kusterExplicitBehavioralConsistency2003}
J.~K{\"u}ster and J.~Stehr, ``Towards explicit behavioral consistency concepts
  in the {{UML}},'' in \emph{Proceedings of 2nd {{ICSE}} Workshop on Scenarios
  and State Machines: {{Models}}, Algorithms, and Tools (Portland, {{USA}})},
  2003.

\bibitem{yaoConsistencyCheckingUML2006}
S.~Yao and S.~Shatz, ``\BIBforeignlanguage{en}{Consistency {{Checking}} of
  {{UML Dynamic Models Based}} on {{Petri Net Techniques}}},'' in
  \emph{\BIBforeignlanguage{en}{2006 15th {{International Conference}} on
  {{Computing}}}}.\hskip 1em plus 0.5em minus 0.4em\relax {Mexico city,
  Mexico}: {IEEE}, Nov. 2006, pp. 289--297.

\bibitem{cunhaFormalVerificationUML2011}
E.~Cunha, M.~Custodio, H.~Rocha, and R.~Barreto,
  ``\BIBforeignlanguage{en}{Formal {{Verification}} of {{UML Sequence
  Diagrams}} in the {{Embedded Systems Context}}},'' in
  \emph{\BIBforeignlanguage{en}{2011 {{Brazilian Symposium}} on {{Computing
  System Engineering}}}}.\hskip 1em plus 0.5em minus 0.4em\relax
  {Florianopolis, Brazil}: {IEEE}, Nov. 2011, pp. 39--45.

\bibitem{thierry-miegUMLBehavioralConsistency2008}
Y.~{Thierry-Mieg} and L.-M. Hillah, ``{{UML}} behavioral consistency checking
  using instantiable {{Petri}} nets,'' \emph{Innovations in Systems and
  Software Engineering}, vol.~4, no.~3, pp. 293--300, Oct. 2008.

\bibitem{ekerTamingHeterogeneityPtolemy2003}
J.~Eker, J.~Janneck, E.~Lee, {Jie Liu}, {Xiaojun Liu}, J.~Ludvig,
  S.~Neuendorffer, S.~Sachs, and {Yuhong Xiong},
  ``\BIBforeignlanguage{en}{Taming heterogeneity - the {{Ptolemy}} approach},''
  \emph{\BIBforeignlanguage{en}{Proceedings of the IEEE}}, vol.~91, no.~1, pp.
  127--144, Jan. 2003.

\bibitem{deantoniModelingBehavioralSemantics2016}
J.~Deantoni, ``\BIBforeignlanguage{en}{Modeling the {{Behavioral Semantics}} of
  {{Heterogeneous Languages}} and their {{Coordination}}},'' in
  \emph{\BIBforeignlanguage{en}{2016 {{Architecture}}-{{Centric Virtual
  Integration}} ({{ACVI}})}}.\hskip 1em plus 0.5em minus 0.4em\relax {Venice,
  Italy}: {IEEE}, Apr. 2016, pp. 12--18.

\bibitem{tripakisModularFormalSemantics2013}
S.~Tripakis, C.~Stergiou, C.~Shaver, and E.~A. Lee, ``\BIBforeignlanguage{en}{A
  modular formal semantics for {{Ptolemy}}},''
  \emph{\BIBforeignlanguage{en}{Mathematical Structures in Computer Science}},
  vol.~23, no.~4, pp. 834--881, Aug. 2013.

\bibitem{elhamlaouiAlignmentViewpointHeterogeneous2016}
M.~El~Hamlaoui, B.~Coulette, S.~Ebersold, S.~Bennani, M.~Nassar, A.~Anwar,
  A.~Beugnard, J.-C. Bach, Y.~Jamoussi, and H.~N. Tran, ``Alignment of
  viewpoint heterogeneous design models: {{Emergency}} department case study,''
  in \emph{4th International Workshop on the Globalization of Modeling
  Languages ({{GEMOC}} 2016) Co-Located with {{ACM}}/{{IEEE MODELS}} 2016},
  {Saint-Malo, France}, Oct. 2016, pp. pp. 18--27.

\bibitem{kuskeFormalSemanticsUML2001}
S.~Kuske, ``A formal semantics of {{UML}} state machines based on structured
  graph transformation,'' in \emph{{$\ll$}{{UML}}{$\gg$} 2001 \textemdash{}
  {{The}} Unified Modeling Language. {{Modeling}} Languages, Concepts, and
  Tools}, M.~Gogolla and C.~Kobryn, Eds.\hskip 1em plus 0.5em minus 0.4em\relax
  {Berlin, Heidelberg}: {Springer Berlin Heidelberg}, 2001, pp. 241--256.

\bibitem{varroFormalSemanticsUML2002}
D.~Varr{\'o}, ``A formal semantics of {{UML}} statecharts by model transition
  systems,'' in \emph{Graph Transformation}, A.~Corradini, H.~Ehrig, H.~J.
  Kreowski, and G.~Rozenberg, Eds.\hskip 1em plus 0.5em minus 0.4em\relax
  {Berlin, Heidelberg}: {Springer Berlin Heidelberg}, 2002, pp. 378--392.

\bibitem{ehrigGraphGrammarsPetri2004}
H.~Ehrig and J.~Padberg, ``Graph {{Grammars}} and {{Petri Net
  Transformations}},'' in \emph{Lectures on {{Concurrency}} and {{Petri
  Nets}}}, T.~Kanade, J.~Kittler, J.~M. Kleinberg, F.~Mattern, J.~C. Mitchell,
  M.~Naor, O.~Nierstrasz, C.~Pandu~Rangan, B.~Steffen, M.~Sudan,
  D.~Terzopoulos, D.~Tygar, M.~Y. Vardi, G.~Weikum, J.~Desel, W.~Reisig, and
  G.~Rozenberg, Eds.\hskip 1em plus 0.5em minus 0.4em\relax {Berlin,
  Heidelberg}: {Springer Berlin Heidelberg}, 2004, vol. 3098, pp. 496--536.

\bibitem{gadducciGraphRewritingPcalculus2007}
F.~GADDUCCI, ``Graph rewriting for the {$\pi$}-calculus,'' \emph{Mathematical
  Structures in Computer Science}, vol.~17, no.~3, pp. 407--437, 2007.

\bibitem{rutleMetamodellingApproachBehavioural2012}
A.~Rutle, W.~MacCaull, H.~Wang, and Y.~Lamo, ``\BIBforeignlanguage{en}{A
  metamodelling approach to behavioural modelling},'' in
  \emph{\BIBforeignlanguage{en}{Proceedings of the {{Fourth Workshop}} on
  {{Behaviour Modelling}} - {{Foundations}} and {{Applications}} -
  {{BM}}-{{FA}} '12}}.\hskip 1em plus 0.5em minus 0.4em\relax {Kgs. Lyngby,
  Denmark}: {ACM Press}, 2012, pp. 1--10.

\bibitem{clarkeHandbookModelChecking2018}
E.~M. Clarke, T.~A. Henzinger, H.~Veith, and R.~Bloem, Eds.,
  \emph{\BIBforeignlanguage{en}{Handbook of {{Model Checking}}}}.\hskip 1em
  plus 0.5em minus 0.4em\relax {Cham}: {Springer International Publishing},
  2018.

\bibitem{ghamarianModellingAnalysisUsing2012}
A.~H. Ghamarian, M.~{de Mol}, A.~Rensink, E.~Zambon, and M.~Zimakova,
  ``\BIBforeignlanguage{en}{Modelling and analysis using {{GROOVE}}},''
  \emph{\BIBforeignlanguage{en}{International Journal on Software Tools for
  Technology Transfer}}, vol.~14, no.~1, pp. 15--40, Feb. 2012.

\bibitem{rensinkGROOVESimulatorTool2004}
A.~Rensink, ``\BIBforeignlanguage{en}{The {{GROOVE Simulator}}: {{A Tool}} for
  {{State Space Generation}}},'' in \emph{\BIBforeignlanguage{en}{Applications
  of {{Graph Transformations}} with {{Industrial Relevance}}}}, T.~Kanade,
  J.~Kittler, J.~M. Kleinberg, F.~Mattern, J.~C. Mitchell, M.~Naor,
  O.~Nierstrasz, C.~Pandu~Rangan, B.~Steffen, M.~Sudan, D.~Terzopoulos,
  D.~Tygar, M.~Y. Vardi, G.~Weikum, J.~L. Pfaltz, M.~Nagl, and B.~B{\"o}hlen,
  Eds.\hskip 1em plus 0.5em minus 0.4em\relax {Berlin, Heidelberg}: {Springer
  Berlin Heidelberg}, 2004, vol. 3062, pp. 479--485.

\bibitem{costaVerigraphSystemSpecification2016}
A.~Costa, J.~Bezerra, G.~Azzi, L.~Rodrigues, T.~R. Becker, R.~G. Herdt, and
  R.~Machado, ``Verigraph: {{A}} system for specification and analysis of graph
  grammars,'' in \emph{Formal Methods: {{Foundations}} and Applications},
  L.~Ribeiro and T.~Lecomte, Eds.\hskip 1em plus 0.5em minus 0.4em\relax
  {Cham}: {Springer International Publishing}, 2016, pp. 78--94.

\bibitem{kunzeBehaviouralModelsModelling2016}
M.~Kunze and M.~Weske, \emph{\BIBforeignlanguage{en}{Behavioural Models: From
  Modelling Finite Automata to Analysing Business Processes}}.\hskip 1em plus
  0.5em minus 0.4em\relax {Cham}: {Springer}, 2016.

\end{thebibliography}

\end{document}